\documentclass{article}

\usepackage{arxiv}

\usepackage[utf8]{inputenc} 
\usepackage[T1]{fontenc}    
\usepackage{hyperref}       
\usepackage{url}            
\usepackage{booktabs}       
\usepackage{amsfonts}       
\usepackage{nicefrac}       
\usepackage{microtype}      
\usepackage{lipsum}

\usepackage{amsmath, setspace,geometry,color}
\usepackage{graphicx}
\usepackage{xcolor}
\usepackage[ruled, boxed, vlined]{algorithm2e}
\usepackage{algorithmic}
\usepackage{multirow}
\usepackage[caption=false,font=footnotesize]{subfig}
\usepackage{url}

\newtheorem{definition}{Definition}

\title{Early Response Assessment in Lung Cancer Patients using Spatio-temporal CBCT Images}

\author{
  Bijju Kranthi Veduruparthi\textsuperscript{*}, Jayanta Mukherjee, Partha Pratim Das\\
  Department of Computer Science and Engineering\\
  Indian Institute of Technology Kharagpur, India\\
  \textsuperscript{*}\texttt{bijjuair@gmail.com} \\
   \And
   Mandira Saha, Sanjoy Chatterjee  \\
   Department of Radiation Oncology\\
   Tata Medical Center, Kolkata, India\\
   \And
   Raj Kumar Shrimali \\
   Arden Cancer Centre\\
   University Hospitals Coventry and Warwickshire\\
   NHS Trust, Coventry, United Kingdom\\
   \And
   Soumendranath Ray\\
   Department of Nuclear Medicine\\
   Tata Medical Center, Kolkata, India\\
   \And
   Sriram Prasath\\
   Department of Medical Physics\\
   Tata Medical Center, Kolkata, India\\
}

\begin{document}
\maketitle

\begin{abstract}
  We report a model to predict patient's radiological response to curative radiation therapy (RT) for non-small-cell lung cancer (NSCLC).
  Cone-Beam Computed Tomography images acquired weekly during the six-week course of RT were contoured with the Gross Tumor Volume (GTV) by senior radiation oncologists for 53 patients (7 images per patient).
  Deformable registration of the images yielded six deformation fields for each pair of consecutive images per patient.
  Jacobian of a field provides a measure of local expansion/contraction and is used in our model.
  Delineations were compared post-registration to compute unchanged ($U$), newly grown ($G$), and reduced ($R$) regions within GTV. 
  The mean Jacobian of these regions $\mu_U$, $\mu_G$ and $\mu_R$ are statistically compared and a response assessment model is proposed.
  A good response is hypothesized if $\mu_R < 1.0$, $\mu_R < \mu_U$, and $\mu_G < \mu_U$.
  For early prediction of post-treatment response, first, three weeks' images are used.
  Our model predicted clinical response with a precision of $74\%$.
  Using reduction in CT numbers (CTN) and percentage GTV reduction as features in logistic regression, yielded an area-under-curve of 0.65 with p=0.005.
  Combining logistic regression model with the proposed hypothesis yielded an odds ratio of 20.0 (p=0.0).
  \end{abstract}

\keywords{NSCLC \and CBCT \and RECIST \and Jacobian \and Classification \and Image Registration}

\section{Introduction}
 Non-small cell lung cancer (NSCLC) accounts for three-quarters of all lung cancer cases. Curative radical radiotherapy is the treatment of choice for most patients without any distant metastases \cite{sause2000final}. The radiotherapy schedule used is typically fractionated, using thirty fractions, each of 2 Gy, delivered over 6 weeks to a total dose of 60 Gy.
  
  Radiation therapy planning requires identifying the tumor based on a pre-treatment planning computed tomography (CT) taken with the patient in a treatment position followed by localization of the tumor by the radiation oncologist. During the process, nearby organs at risk are also identified and delineated to ensure that radiation induced damage is avoided or minimized.
  Modern radiation therapy machines acquire cone beam CT (CBCT) images on a daily or a weekly basis prior for delivery of the radiation to identify and minimize any positional changes in the anatomy and tumor.
  Such image-guided curative radiation therapy (IGRT) is considered a standard of care for locally advanced lung cancers.
  Tumor response on CT scans, three months post therapy completion, was assessed by the clinicians using Radiology Response Evaluation Criteria in Solid Tumors (RECIST) \cite{eisenhauer2009new} criteria. Based on the tumor size and metastasis, the patient is classified into one the categories as Complete Response (CR), Partial Response (PR), Stable Disease (SD), or Progressive Disease (PD).
  A responsive patient has tumor regression which corresponds to either CR or PR categories. Similarly, a non-responsive patient has tumor growth which corresponds to SD or PD categories.
  In this work, we developed a model that could identify post-therapy treatment response using Jacobian statistics.
  Importantly, we explored whether it is possible to predict response post-treatment completion, following analysis of the CBCT images captured during the first three weeks of radiation course.
  In our work, a responsive patient has a RECIST of either CR or PR.
  Similarly, non-responsive patients have a RECIST of either SD or PD. 
  
  \subsection{Related Works}
  
  Radiomics is a branch of medical science that aims to extract knowledge from medical images of patients for their clinical diagnosis, treatment, and prognosis \cite{gillies2015radiomics}.
  There are several works in the literature addressing clinical diagnosis and treatment using images.
  Prediction of the response of a patient to radiation treatment is done by estimating statistics like Progression Free Survival (PFS), Overall Survival (OS), RECIST criteria, etc.
  Prognosis is an estimate of the effectiveness of the delivered treatment.
  Different machine learning techniques that were applied in cancer prognosis are discussed in the literature \cite{Kourou2015, Lynch2017}.
  Here, the features used for regression or classification are non-image data like age, gender, stage of cancer, histology, and other several diagnostic and clinical findings.
  Radiomic features based on morphology, image statistics, texture, and fractal analysis were used in \cite{lee2017radiomics}.
  Similar studies are \cite{coroller2016radiomic, hunter2016nsclc, birchard2009early, fave2017delta, shi2018radiomics}.
  
  Several approaches to prognosis have been explored using image data. Here, we review a few that are relevant to this work.
  In a recent work \cite{wen2017value}, the reduction in the Computed Tomography Number (CTN) and the Gross Tumor Volume (GTV) over 6 weeks of chemoradiation therapy was collected in 54 NSCLC patients.
  Serial CBCT images of these patients taken during treatment were delineated with the GTV by experienced radiologists. The mean CTN numbers within the GTV were computed.
  A reduction in the mean CTN and GTV were computed for a period of 6 weeks.
  It was observed that there is a different pattern in CTN and GTV reduction in responsive and non-responsive lung cancer patients.
  The difference in the mean CTN in the first and the last week was statistically significant.
  The GTV reduction in responsive patients was higher than that of non-responsive patients.
  A logistic regression classifier was developed using the mean CTN and GTV reduction as features.
  The sensitivity, specificity, positive predictive and negative predictive values reported on their dataset were 58.8\%, 86.4\%, 87.16\%, and 57.17\%, respectively.
  Early response to treatment was possible by considering the 3 weeks CTN and GTV reduction values in the classifier.
  In this case, the sensitivity, specificity, positive predictive and negative predictive values reported on their dataset were 58.8\%, 90.9\%, 90.6\%, and 58.3\%, respectively.
  However, a large collection of patients is required to further validate the hypothesis.
  \begin{figure*}[!hb]
  \centering
  \includegraphics{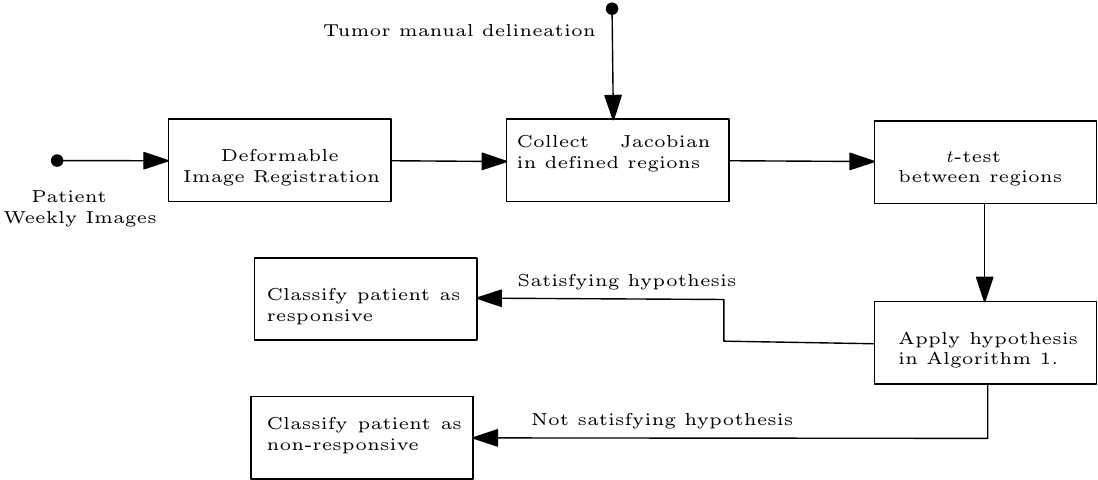}
  \caption{The overall picture of classifying each patient in the population and test data is shown here as a block diagram.}
  \label{fig:BlockDiagram}
  \end{figure*}
  
  In one approach\cite{Jabbour2015},  the manually delineated GTV in serial CBCT scans of 38 patients were taken.
  The reduction in GTV was correlated with survival using a univariate Cox proportional hazards regression model.
  Using this model, it was predicted that every 10\% reduction in GTV led to 44\% reduction in the risk of death.
  In another work\cite{Brink2014}, serial CBCT scans of 99 patients with NSCLC were collected.
  The Jacobian determinant along with the GTV at one-third of treatment and two-thirds of treatment were used to predict the GTV at the end of treatment.
  Early regression of tumor volume was computed by fitting an exponential function of time to the GTV collected before one-third or two-thirds of treatment.
  This work aimed to predict the GTV at the end of treatment but did not attempt to predict the response of a patient to treatment.
  In another study \cite{Hou2015}, GTV delineated in CT were propagated to serial CBCTs in 30 head and neck cancer patients.
  Regression of the GTV after registering the CT onto the CBCTs was computed.
  A correlation between RECIST diameter measurements and GTV regression was computed.
  It was observed that the GTV regression did not always correlate with RECIST diameter regression.
  However, the GTV regression could differentiate between complete response and non-complete response patients.
  In our dataset, extra CT scans were not performed to delineate serial GTVs rather GTV was contoured on the CBCTs performed weekly for image guidance during treatment. 
  These data sets were analyzed for predicting response to therapy and to predict response 3 months post-therapy.
  
  Computational Anatomy \cite{Miller2009} using deformable templates that exhibit diffeomorphism, a bijection between anatomical coordinates was created in our study.  Using Jacobian maps we identified the direction of growth, stable areas and regression in the various spatio-temporal directions of the GTV. 
  The deformable template that exhibits diffeomorphism, a bijection between anatomical coordinates is sometimes used in such a study. 
  Landmarks \cite{Miller2004} around the area of interest were used to track changes, where the Jacobian maps showed the direction of growth in ventral and dorsal parts of the hippocampus of a mouse. Deformation fields obtained after deformable image registration 
  were earlier used for characterizing different regions of the lung \cite{Amelon2010}, based on the motion properties like directional change, volume change and nature of change.
  In another study \cite{cao2012tracking}, the deformation fields were used to track changes in tissue volumes. 
  Different regions of the lung were segregated into blocks, and average Jacobian in these regions was observed to report lung activity and for tracking the change in volume of tissues.
  In another work, anatomical changes in the brain \cite{lorenzi2013sparse} were observed to be different across spatial scales using the Helmholtz decomposition of the deformation field. 
  A difference of Gaussians (DOG) operator was applied to the irrotational component of the decomposition to identify the areas of maximal volume change. 
  However, this method was applied only to brain MR images. 
  
  \subsection{Overview of our method}
  Earlier methods are discussed previously made use of PET and CT images for computing radiomic features, and response prediction.
  The proposed method uses spatio-temporal CBCTs as they are low dose images used for regular scanning of a patient during chemoradiation.
  In our method, we use Jacobians of voxels within the GTV to predict the response of a patient.
  An overview of the proposed workflow is shown in Fig.~\ref{fig:BlockDiagram}.
  Weekly images are registered and Jacobians are computed in the proposed regions.
  Statistical tests are used to design a hypothesis for classification of responsive and non-responsive patients.
  
  \section{Method}
  The datasets for this study were obtained from a larger cohort of 251 patients with lung cancer, who were treated using radical radiotherapy with
  curative intent \cite{shrimali2018impact}. The survival outcomes for this group have been reported with subgroup analyses for sequential chemoradiation,
  concurrent chemoradiation and radiotherapy alone \cite{shrimali2018impact}. Analyses have also been reported for patients treated using 3-dimensional conformal
  radiotherapy (3D-CRT) versus volumetric modulated radiotherapy (VMAT) and patients with larger target volumes (PTV $> 500ml$) versus smaller
  target volumes (PTV $< 500ml$) \cite{shrimali2018impact}. It has been shown that the treatment protocol \cite{shrimali2018impact} has resulted in outcomes that are at par
  with published literature from large multi-centre phase-III clinical trials (RTOG 0617) as well as retrospective data from the UK \cite{iqbal2019hypofractionated, bradley2015standard, chun2017impact, arunsingh2018survival}.
  From this dataset of radically non-small cell lung cancer (NSCLC) patients treated in a tertiary care center, fifty-three random patients who received conventionally fractionated RT over 6 weeks were selected. All of these patients had completed their routine treatment using concurrent chemoradiation (60Gy in 30 fractions over six weeks, with concurrent cisplatin and etoposide). Patients treated with hyperfractionated accelerated RT and hypofractionated accelerated RT were excluded from this study in order to achieve uniformity in the selected group and allow us to test the three-week prediction model as proposed above.
  The Cancer Hospital - Institutional Review Board discussed our study using CBCT from archived data of patients and did not recommend informed consent from patients for this study.
  All protocols and methods in the study were in agreement with the guidelines and regulations.
  
  We performed a statistical analysis of deformation fields obtained by registering CBCT images of lung cancer patients who underwent radiotherapy.
  \begin{figure*}[!hb]
  \centering
  \subfloat[]{\includegraphics[width=0.59\textwidth]{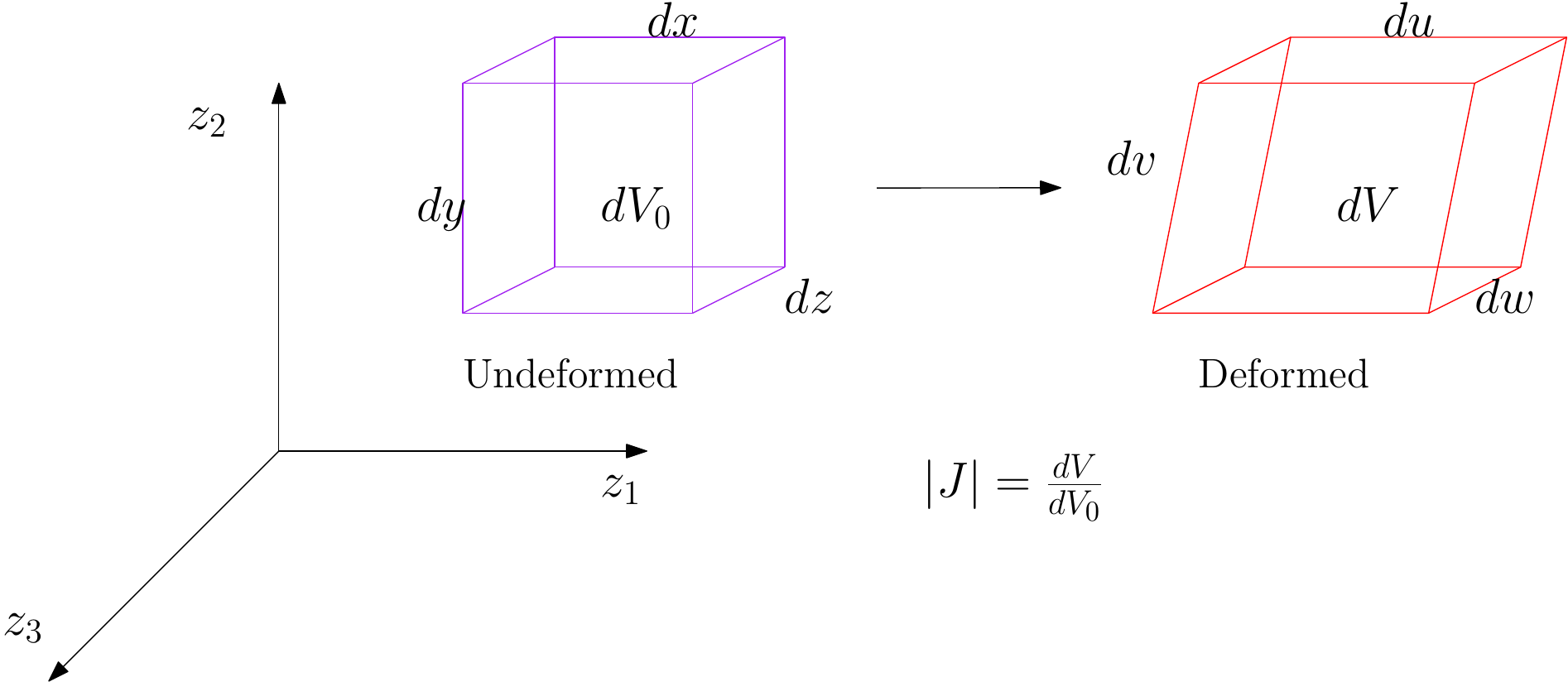}
  \label{fig:JacobianVolume}}
  \subfloat[]{\includegraphics[width=0.39\textwidth]{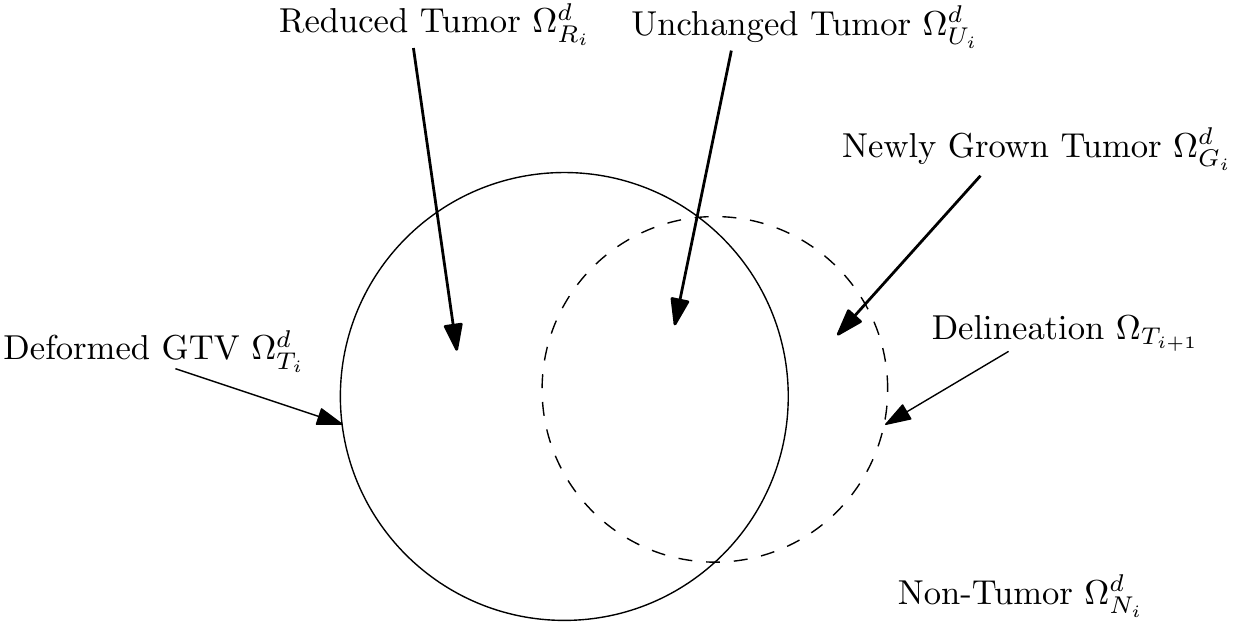}
  \label{fig:GrowthDecayModel}}
  \caption{(a) Jacobian computes the ratio of the deformed volume to the reference undeformed volume. (b) Different regions (R,G,U) of the tumor for two overlapping synthetic tumors. Solid line shows source tumor delineated volume. Dotted line shows the target volume.}
  \label{fig:JacobianModel}
  \end{figure*}
  
  The delineated tumor voxels were categorized into different regions. We studied the Jacobian statistic on the deformation fields in these regions, and study how it varies across patients. Patients satisfying our proposed hypothesis were found to have a better response to treatment. The results of the classification suggest that early prediction of responses to radiotherapeutic treatment is possible.
  
  This work involved the statistical analysis of deformation fields obtained by registering serial on-treatment CBCT images of lung cancer patients who underwent radiotherapy. 
  The serial GTV delineations were used to categorize the tumor into different regions i.e. reduced (R), newly-grown (G) and unchanged (U) depending on the respective areas identified in the serial imaging after registration. We studied the Jacobian statistics on the deformation fields in these regions.
  We explored if using the Jacobian statistics regression could be as accurate and precise as RECIST reports reporting response. 
  Jacobian of each voxel within the Volume of interest (VOI) of serial CBCT from every patient was analyzed.
  The VOI included the entire GTV and $\pm20$ voxels around it.
  Using the Jacobian collected in this VOI, and categorized into the 4 regions (R, G, U, and N), a hypothesis was proposed (Algorithm 1) to differentiate a responsive from a non-responsive patient. 
  
  \subsection{Image Registration}\label{img_reg}
  The image registration process consists of rigid registration of a pair of CBCT images in the first stage, followed by their deformable image registration in the next stage.
  The rigid registration is implemented using the built-in algorithm in the Insight ToolKit (ITK) \cite{yoo2002engineering_itk}, which uses the concept of mutual information \cite{mattes2003pet} to find the optimal alignment.
  A 3-level multiresolution scheme is employed here to first register the images at $\frac{1}{4}^{th}$ the original resolution.
  The registration parameters obtained at this resolution are used to initialize the registration parameters at $\frac{1}{2}$ the original resolution.
  Finally, using them the registration parameters at the full resolution are obtained.
  The registration parameters consist of 3 translation and 3 rotation parameters with respect to the center of the image.
  As a preprocessing step, the images were smoothed using Gaussian smoothing with a standard deviation of 2.0, 1.0 at the resolutions of $\frac{1}{4}$, $\frac{1}{2}$ of the original resolution.
  The probability density functions of the pair of CBCT images were computed using the Parzen window technique, with a bin size of 50.
  We used the gradient descent optimization scheme to find the optimal parameters with the step size of 1.0, limiting to a maximum of 100 iterations.
  The second stage of image registration involves deformable image registration.
  We have used the non-linear registration technique, where the symmetric Local Correlation Coefficient (LCC) \cite{Lorenzi2013} is used as the similarity measure. 
  The output of the first stage of registration is used as the transformed input image to register onto the next CBCT image.

  Given two images $S$, $T \in \mathbb{R}^3$, the goal of image registration is to find a transformation
  $\vec{g} : \mathbb{R}^3 \mapsto \mathbb{R}^3$ that maps/aligns $S$ onto $T$.
  The approach of computing both the forward (registering $S$ to $T$) and reverse (registering $T$ to $S$) transformations is termed as bidirectional symmetric registration. This process involves the computation of inverse of the deformation field.
  
  The Jacobian of a deformation field $\phi$ is defined as:
  \begin{equation}
  \label{eq:JacobianDet}
  J(\vec{\phi}(\vec{z})) = \begin{bmatrix} \frac{\partial \phi_1(\vec{z})}{\partial z_1} & \frac{\partial \phi_1(\vec{z})}{\partial z_2}
    & \frac{\partial \phi_1(\vec{z})}{\partial z_3} 
    & \\ \frac{\partial \phi_2(\vec{z})}{\partial z_1} & \frac{\partial \phi_2(\vec{z})}{\partial z_2} & \frac{\partial \phi_2(\vec{z})}{\partial z_3} 
    & \\ \frac{\partial \phi_3(\vec{z})}{\partial z_1} & \frac{\partial \phi_3(\vec{z})}{\partial z_2} & \frac{\partial \phi_3(\vec{z})}{\partial z_3}
  \end{bmatrix}
  \end{equation}

  \begin{figure*}[!ht]
  \centering
  \subfloat[]{\includegraphics[width=0.33\textwidth,height=0.33\textwidth]{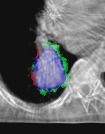}
  \label{fig:Concost}}
  \subfloat[]{\includegraphics[width=0.33\textwidth,height=0.33\textwidth]{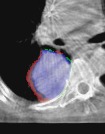}
  \label{fig:RegCost}}
  \subfloat[]{\includegraphics[width=0.33\textwidth,height=0.33\textwidth]{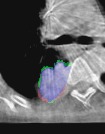}
  \label{fig:SimCost}}
  \caption{The images (a), (b), (c) show several slices of the CBCT images along with their delineations highlighted for weeks $i$ and $i+1$. The green region represents the reduced region ($R_i$), blue region represents the unchanged region ($U_i$), and red region represents the newly grown region ($G_i$), respectively.}
  \label{fig:DispImage}
  \end{figure*}

  Deformable registration of a pair of images yielded a single deformation field.
  Each pair of consecutive CBCT image was registered onto the next weekly CBCT image except the last image.
  Therefore, the 7 CBCT images yielded 6 deformation fields.
  The deformation index for interpreting the information in a deformation field was the determinant of Jacobian, commonly referred to in the literature as simply Jacobian.
  The Jacobian $(J)$ measured the local volume change with respect to that of a unit cube.
  This can be visualized in Fig.~\ref{fig:JacobianVolume}.
  When $J$ varies between $0$ and $\infty$, the deformation field is a diffeomorphic transformation.
  When $J = 1$, it means there is no change in volume.
  A value of $J < 1$ denoted, net contraction; and $J > 1$ denoted, net expansion.
  $J$ can also be negative, implying that there was either folding/cutting of the voxel grid, resulting in a non-diffeomorphic deformation field.
  Clearly, the negative $J$ is not desirable as there is no such folding/cutting physical phenomenon in the considered images.
  
  \subsection{Categorization of Regions}\label{regions}
  The deformation field obtained by registering the weekly CBCT images can be correspondingly used to deform the GTV in each image.
  The two consecutive pair of GTVs, one comprising of the deformed GTV, and the other consisting of the next week's undeformed GTV can be overlapped.
  
  Let us define $\Omega_T$ as the support space where the tumor is active, which corresponds to locations within the GTV of a patient, and $\Omega^{d}_{T_i}$ represent its deformed space at the $i^{th}$ week of treatment.
  Using this definition, the different regions of the GTV can be defined and categorized into the following:
  \begin{definition}
  \begin{itemize}
  \item []
  \item \textbf{Deformed Tumor Region} in week $i$ ($\Omega^{d}_{T_i}$): Set of tumor voxels in the deformed GTV of week $i$.
  \item \textbf{Tumor Region} in week $i$ ($\Omega_{T_i}$): Set of tumor voxels in the undeformed GTV of week $i$.
  \item \textbf{Non-Tumor Region} in week $i$ ($\Omega^{d}_{N_i}$): Set of non-tumor voxels computed as negation of the set $\Omega^{d}_{T_i} \cup \Omega_{T_{i+1}}$.
  \item \textbf{Unchanged Region} in week $i$ ($\Omega^{d}_{U_i}$): The set of tumor voxels that remain unchanged i.e, $\Omega^{d}_{T_i} \cap \Omega_{T_{i+1}}$.
  \item \textbf{Reduced Region} in week $i$ ($\Omega^{d}_{R_i}$): The set of tumor voxels that is defined as, $\Omega^{d}_{T_i} \setminus \Omega_{T_{i+1}}$.
  \item \textbf{Newly grown Region} in week $i$ ($\Omega^{d}_{G_i}$): The set of tumor voxels that is defined as, $\Omega_{T_{i+1}} \setminus \Omega^{d}_{T_i}$.
  \end{itemize}
  \end{definition}
  
  The categorization can be visualized in Fig.~\ref{fig:GrowthDecayModel}.
  The intuitive understanding of these regions is as follows:
  $R$ represents the region that was present in the initial GTV, but not present in next week's GTV.
  Similarly, $G$ represents the region that was not present in the initial GTV, but present in next week's GTV.
  Finally, $U$ represents the region that was present in both the initial GTV and next week's GTV.
  However, $N$ region was neither present in the initial GTV not in the next week's GTV.
  Applying the above process on real images, several slices are shown in Fig.~\ref{fig:DispImage}. We study the deformation fields in these regions.
  The mean Jacobian ($\mu$) in these different areas of the 3D image is computed, and statistical tests are used to test a hypothesis for classifying patients.
  Mathematically, they can be formulated using the following expressions:
  \begin{equation}
  \begin{aligned}
   \mu_R = \frac {1} { | \Omega^{d}_{R_i} | } \sum_{\vec{z} \in \Omega^{d}_{R_i}} J(\vec{g}(\vec{z})) \\
   \mu_U = \frac {1} { | \Omega^{d}_{U_i} | } \sum_{\vec{z} \in \Omega^{d}_{U_i}} J(\vec{g}(\vec{z})) \\
   \mu_G = \frac {1} { | \Omega^{d}_{G_i} | } \sum_{\vec{z} \in \Omega^{d}_{G_i}} J(\vec{g}(\vec{z})) 
  \end{aligned}
  \end{equation}
  
  \subsection{Ordering hypothesis}
  Let us denote the mean($\mu$) and standard deviation($\sigma$) of the Jacobians in the regions with the respective subscripts, where $T$ corresponds to tumor regions, $N$ corresponds to non-tumor regions, $U$ corresponds to unchanged regions, $R$ corresponds to reduced regions, and $G$ corresponds to newly grown regions.
  
  We performed the two-sided \emph{t}-test for each of the patients to record the ordering of the means in different regions.
  The ordering from \emph{t}-test was used in the hypothesis for classifying patients into responsive and non-responsive categories.
  It was designed based on the following intuitive logic: 
  A reduction in GTV was used to categorize patients using the RECIST criteria. 
  It was expected that a responsive patient had better GTV reduction compared to a non-responsive patient.
  Jacobian gives reduction or expansion at each voxel.
  Therefore, a patient with a good response should give the sum of Jacobians at all voxels within the GTV to reduce.
  By definition, the $R$ region should result in a net reduction of the volume. Hence, the mean of Jacobian $\mu_R$ should be less than $1.0$, and it should also be lower than the unchanging region $U$.
  By definition, we know that the expansion in $G$ and $U$ should be greater than that of $R$, because of more tumorous activity in these regions.
  However, a smaller $G$ means lower activity in this region. Therefore, we hypothesize that $\mu_{G} \leq \mu_{U}$. 
  
  \begin{algorithm}[!htbp]
  \label{algo:ordering_algo}
  \SetAlgoLined
  \KwData{\emph{t}-test ordering result and Jacobian means in different regions for every patient}
  \KwResult{Binary classification of patient response for each patient }
    \eIf{$\mu_R \leq 1.0$ and $\mu_R \leq \mu_U$ and $\mu_G \leq \mu_U$}{
     Classify patient as responsive\;
     }{
     Classify patient as non-responsive\;
    }
   \caption{Ordering hypothesis}
  \end{algorithm}
  
  The hypothesis is formalized in Algorithm.1. 
  The block diagram of the classification process is shown in Fig.~\ref{fig:BlockDiagram}.
  The proposed hypothesis is an unsupervised method for classification and is therefore independent of training data.
  
  \subsection{Prediction of Response}
  We followed the method for predicting responsive and non-responsive outcomes as proposed in \cite{wen2017value} on our data.
  Here, binary logistic regression was used on the entire dataset to classify patients into responsive and non-responsive patients.
  However, it is observed in the literature that there is a high chance for overfitting, as experiments as reported in that work, did not consider training samples and test samples separately.
  The features used were the reduction in Computed Tomography Number (CTN) and the percentage reduction in Gross Tumor Volume (GTV), across the period of treatment.
  In our data also, CTN and GTV were collected in serial CBCT scans during a median period of 6 weeks of treatment.
  The mean CTV reduction ($\mu^{R}_{CTN_i}$) at end of week $i$ of the treatment is defined as follows:
  $\mu^{R}_{CTN_i} = \mu_{CTN_i} - \mu_{CTN_0}$; where $\mu_{CTN_i}$ is the mean CTN within GTV at week i.
  Similarly, the percentage GTV reduction ($GTV^R_i$) at the end of week $i$ can be defined as:
  $GTV^R_i = \frac{GTV_i - GTV_0}{GTV_0} \times 100$; where $GTV_i$ is the GTV at week i.
  Here, $i$ ranges from 1 to 6, used to index the number of weeks of treatment.

  In our proposed method we used a hypothesis using the statistics of Jacobian computed from the serially registered data set, as discussed above.
  We compared our method with the work proposed by Wen~\cite{wen2017value}.
  Additionally, we combined both the method of \cite{wen2017value} with our proposed hypothesis to present a more robust prediction model.
  To prevent overfitting, we use the k-fold cross-validation technique, wherein a small percentage of the training data is kept as test data.
  We have used three-fold cross-validation of the logistic regression model and taken a consensus result while predicting RECIST criteria for responsive and non-responsive patients.
  We predicted the patient as responsive if there was a consensus using the logistic regression model and the proposed hypothesis also predicted a RECIST corresponding to a good response (PR, CR).
  Similarly, we predicted the patient as non-responsive, if there was a consensus using the logistic regression model and the proposed hypothesis also predicted a RECIST corresponding to a bad response (SD, PD).
  In all other cases, there was no consensus, the classifier was unable to take a decision, and there was no conclusion taken.

  \section{Results}
  \begin{table}[]
  \centering
  \caption{Patient demographics and response characteristics.}
  \label{tab:patient_demographics}
  \begin{tabular}{|c|c|}
  \hline
  Characteristics & Category: N                                                                                                                                           \\ \hline
  Stage           & \begin{tabular}[c]{@{}c@{}}IIA: 1\\ IIB: 4 \\ IIIA: 34\\ IIIB: 21\end{tabular}                                                                        \\ \hline
  Histology       & \begin{tabular}[c]{@{}c@{}}Adenocarcinoma: 22\\ Squamouscarcinoma: 32\\ Other: 6\end{tabular}                                                         \\ \hline
  Gender          & \begin{tabular}[c]{@{}c@{}}Female: 7\\ Male:53\end{tabular}                                                                                           \\ \hline
  Age             & \begin{tabular}[c]{@{}c@{}}Median: 62 years\\ Range: 33 - 81 years\end{tabular}                                                                       \\ \hline
  RECIST          & \begin{tabular}[c]{@{}c@{}}Complete Response: 3\\ Partial Response: 29\\ Stable Disease: 12\\ Progressive Disease: 9\\ Not Applicable: 7\end{tabular} \\ \hline
  \end{tabular}
  \end{table}
  
  \subsection{Patient Characteristics}
  Our dataset consists of 53 patients, among whom 31 patients had stage IIIA, 18 had stage IIIB, 1 patient had stage IIA and 3 patients had stage IIB cancer.
  47 patients were males and 6 patients were females with a median age of 62 years (range, 33-81 years).
  All patients had a follow-up contrast-enhanced CT scan done 3 months post-therapy completion and RECIST response criteria were used as the reference for therapy response.
  We did not consider patients with metastasis in our dataset.
  Table.~\ref{tab:patient_demographics} summarizes the patient demographics and response characteristics in our dataset.
  
  \subsection{Image Registration}
  The registered images were manually checked for any misalignments. It was noted that the registration was satisfactory in all cases.
  The computed Jacobian map from a deformation field was spatially smooth and had no noticeable noise.
  
  \subsection{CTN, GTV and Jacobian distributions}
  The mean CTN in the first week was -34.15 $\pm$ 80.63 HU, while the mean CTN in the last week was -60.47 $\pm$ 77.83 HU.
  For all patients the mean CTN across all weeks was -46.27 $\pm$ 78.34 HU.
  For responsive patients the mean CTN across all weeks was -57.08 $\pm$ 79.91 HU.
  For non-responsive patients the mean CTN across all weeks was -29.80 $\pm$ 74.74 HU.
  The mean CTN reduction was less in the first week compared to the last week's mean CTN reduction with a p-value of 0.03 in responsive and 0.00 in non-responsive patients.
  
\begin{figure*}[ht]
  \centering
  \subfloat[]{\includegraphics[width=0.49\textwidth]{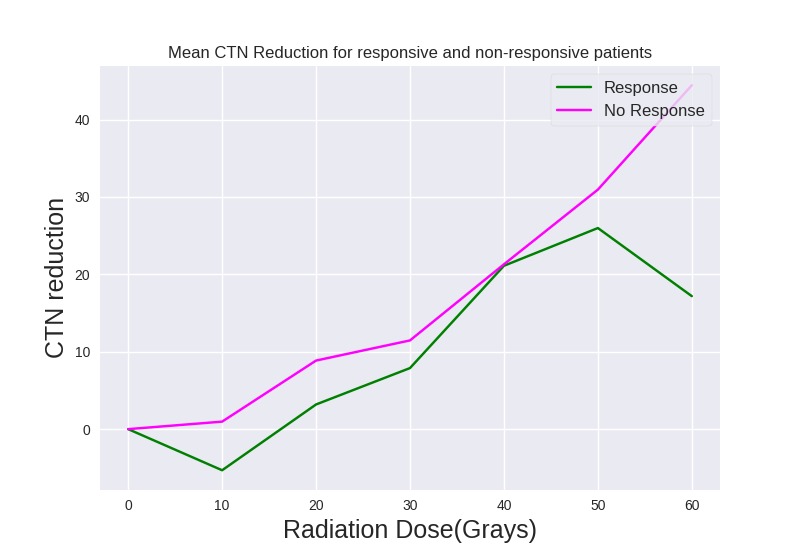}
  \label{fig:CTNDose}}
  \subfloat[]{\includegraphics[width=0.49\textwidth]{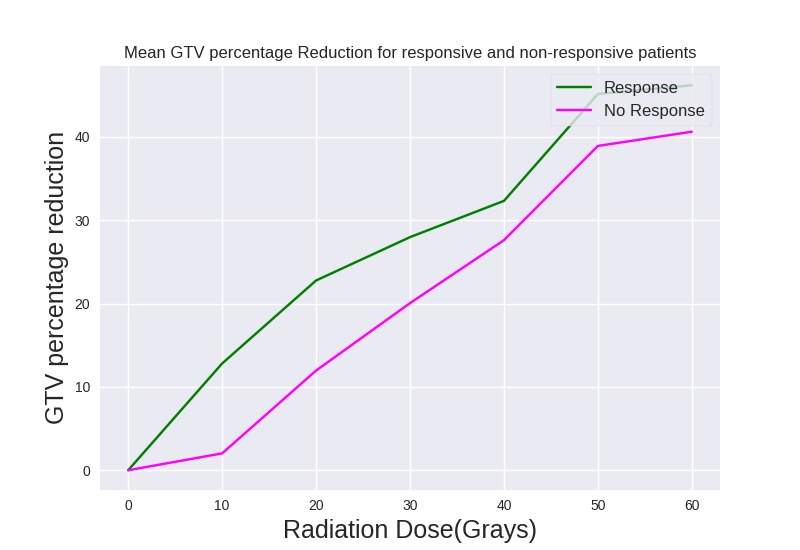}
  \label{fig:GTVDose}}
  \caption{(a) CTN reduction versus Radiation Dose and, (b) GTV percentage reduction versus Radiation Dose, for responsive and non-responsive patients.}
  \label{fig:CTNGTVDose}
  \end{figure*}

  The mean GTV for all patients across all weeks was 129.48 $\pm$ 151.87 $cm^3$.
  For responsive patients, the mean GTV across all weeks was 92.31 $\pm$ 66.73 $cm^3$.
  For non-responsive patients it was 186.13 $\pm$ 217.92 $cm^3$.
  For 36 patients the regression in GTV was greater 30\% from first to last week.
  The mean GTV of all patients reduced from 181.26 $\pm$ 233.09 $cm^3$  in the first week to 91.33 $\pm$ 97.28 $cm^3$ in the last week of treatment. 
  The GTV was less in the first week compared to the last week's GTV with a p-value of 0.00 in both responsive and non-responsive patients.
  The mean CTN and GTV changes with respect to the radiation dose is shown in Fig.~\ref{fig:CTNGTVDose}. 
  
\begin{figure}[ht]
  \centering
  {\includegraphics[width=0.49\textwidth]{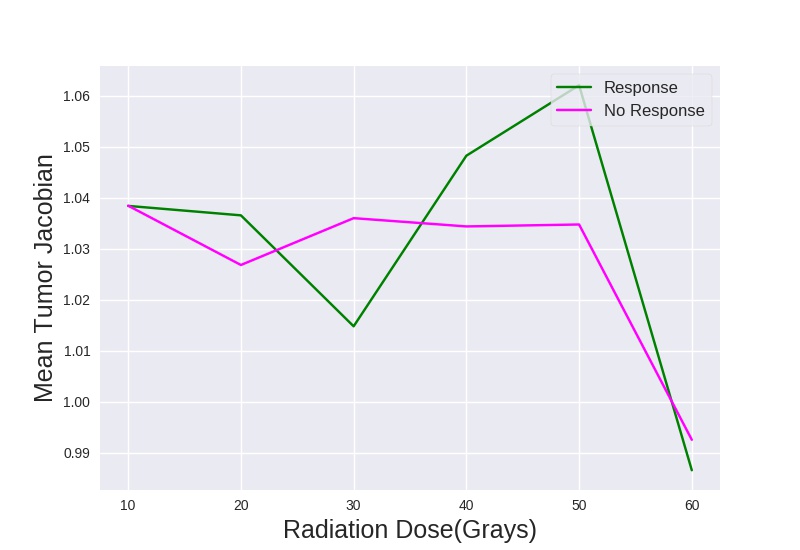}}
  \caption{Jacobian versus Radiation Dose for responsive and non-responsive patients.}
  \label{fig:JacobianDose}
  \end{figure}

  Fig.~\ref{fig:JacobianDose} shows the mean Jacobian for responsive and non-responsive patients over the treatment period.
  The mean Jacobian within the GTV during the first week of treatment was 1.039 $\pm$ 0.079, while in the last week it was 1.112 $\pm$ 0.093.
  In responsive patients, the mean Jacobian in GTV changed from 1.041 $\pm$ 0.075, to 1.106 $\pm$ 0.092.
  While in non-responsive patients the mean Jacobian in GTV changed from 1.036 $\pm$ 0.083, to 1.120 $\pm$ 0.094.
  
  \subsection{Prediction of RECIST}
  Out of the 53 patients in our dataset, 29 patients had PR, 3 had CR, 12 had SD and 9 had PD according to the RECIST criteria.
  
  \subsubsection{Using only Jacobian}
  Statistical tests on the Jacobian are performed for the entire dataset on observing the distribution of J in regions of different categories as described earlier.
  We assume normality of the dataset in all the regions due to the very large number of samples. The number of samples in $R$, $G$, $U$, and $N$ regions are approximately, $1.8$, $1.5$, $3.5$ and $169$ million, respectively.
  
  Using only the first three weeks of Jacobians as data, for responsive patients we found $\mu_R = 0.979 \pm 0.028$, $\mu_G = 1.005 \pm 0.051$, $\mu_U = 1.061 \pm 0.082$, and $\mu_N = 0.985 \pm 0.009$.
  Similarly, for non-responsive patients we found $\mu_R = 1.005 \pm 0.047$, $\mu_G = 1.023 \pm 0.066$, $\mu_U = 1.052 \pm 0.058$, and $\mu_N = 0.985 \pm 0.006$.
  
  Using the proposed hypothesis, 21 of the 32 responsive patients were predicted as responsive,
  while 9 of the 21 non-responsive patients were predicted as non-responsive.
  The results were not statistically significant, where the odds ratio was equal to 1.43 with a p-value of 0.57, implying the result is not statistically significant.
  The sensitivity, specificity, precision, and accuracy are listed in Table.~\ref{tab:summary} (column 2), and were found to be 66\%, 43\%, 64\%, and 57\%, respectively.
  
  However, using only the first three weeks of data provided promising results.
  Using the proposed hypothesis, 20 of the 32 responsive patients were predicted as responsive,
  while 14 of the 21 non-responsive patients were predicted as non-responsive.
  The odds ratio was equal to 3.33 with a p-value of 0.05.
  The sensitivity, specificity, precision, and accuracy are listed in Table.~\ref{tab:summary} (column 3), and were found to be 62\%, 67\%, 74\%, and 64\%, respectively.
  
  This analysis suggests that the hypothesis is satisfied for responsive patients during the first three weeks of the treatment. This early prediction can help doctors take necessary actions for improving the response of the patient to treatment. 
  
  \begin{table*}[]
  \centering
  \caption{Performance of the three prediction methods using full data and early prediction using 3 weeks data.}
  \label{tab:summary}
  \resizebox{1.0\columnwidth}{!}{%
  \begin{tabular}{|c|c|c|c|c|c|c|}
  \hline
          & {\color[HTML]{3166FF} \begin{tabular}[c]{@{}c@{}}Jacobian\\ 6 Weeks\end{tabular}} & {\color[HTML]{3166FF} \begin{tabular}[c]{@{}c@{}}Jacobian\\ 3 Weeks\end{tabular}} & {\color[HTML]{3166FF} \begin{tabular}[c]{@{}c@{}}CTN,GTV\\ 6 Weeks\end{tabular}} & {\color[HTML]{3166FF} \begin{tabular}[c]{@{}c@{}}CTN,GTV\\ 3 Weeks\end{tabular}} & {\color[HTML]{3166FF} \begin{tabular}[c]{@{}c@{}}CTN,GTV, Jacobian\\ 6 Weeks\end{tabular}} & {\color[HTML]{3166FF} \begin{tabular}[c]{@{}c@{}}CTN,GTV, Jacobian\\ 3 Weeks\end{tabular}} \\ \hline
  {\color[HTML]{3166FF} Odds Ratio}  			& 1.43    & 3.33    & 3.5 to 7.23 (4.7)  & 3.5 to 15.5 (9.7)  & 5.0  & 20.0  \\ \hline
  {\color[HTML]{3166FF} p-value}     			& 0.57    & 0.05    & 0.046 to 0.099     & 0.004 to 0.099  & 0.15  & 0.0  \\ \hline
  {\color[HTML]{3166FF} AUC}         			& -       & -       & 0.61  		 & 0.65   	   & -     & -     \\ \hline
  {\color[HTML]{3166FF} Precision \%}			& 64      & 74      & 66.6     		 & 68.9   	   & 71    & 77    \\ \hline
  {\color[HTML]{3166FF} Recall \%}   			& 66      & 62      & 93.75    		 & 96.87  	   & 88    & 95    \\ \hline
  {\color[HTML]{3166FF} Specificity \%}  			& 43      & 67      & 28.57    		 & 33.3   	   & 40    & 50    \\ \hline
  {\color[HTML]{3166FF} Negative Predictive Value \%}	& 45      & 54      & 75   		 & 87.5   	   & 67    & 86    \\ \hline
  {\color[HTML]{3166FF} Accuracy \%} 			& 57      & 64      & 68    		 & 71.7   	   & 70    & 79    \\ \hline
  \end{tabular}
  }
  \end{table*}
  
  \subsubsection{Using CTN and GTV}
  
  CTN reduction and GTV percentage reduction were used as features to classify using a logistic regression model.
  We used a three-fold cross-validation scheme to validate the performance of the classification process.
  We repeated the classification process by shuffling and sampling the data for 50 iterations, in order to look at the variation in the performance.
  The best classifier selected was the one with the maximum AUC.
  In Table.~\ref{tab:summary} (columns 4,5), we list the performance on full data and 3 weeks, respectively.
  We considered only the results where there was a statistical significance of at least 0.1.
  We computed the median odds ratio when using CTN and GTV.
  It was observed that the odds ratio lies between 3.5 and 7.23 with a median odds ratio of 4.7 using the full dataset.
  In the case of three-weeks data, the odds ratio lies between 3.5 and 15.5 with a median odds ratio of 9.7.
  
  \begin{itemize}
  \item All the weeks:
  The maximum area under the curve (AUC) was found to be 0.61 with an odds ratio of 6.0.
  The results corresponding to the maximum AUC are listed in Table.~\ref{tab:summary} (rows 4 - 9, columns 4).
  When classified individually using either of CTN or GTV, gave an average AUC of 0.47 and 0.52, respectively.
  The sensitivity, specificity, precision, and accuracy were found to be 93.75\%, 28.75\%, 66.6\%, and 68\%, respectively, with a p-value of 0.046 in the best case.
  
  \item First 3 weeks: 
  When classified individually using either of CTN or GTV, gave an average AUC of 0.49 and 0.57, respectively.
  The maximum AUC using both CTN and GTV was 0.65 with an odds ratio of 15.5.
  The results corresponding to the maximum AUC are listed in Table.~\ref{tab:summary} (rows 4 - 9, columns 5).
  The sensitivity, specificity, precision, and accuracy were found to be 96.87\%, 33.3\%, 68.9\%, and 71.7\%, respectively, with a p-value of 0.004 in the best case.
  \end{itemize}
  
\begin{figure*}[ht]
  \centering
  \subfloat[]{\includegraphics[width=0.49\textwidth]{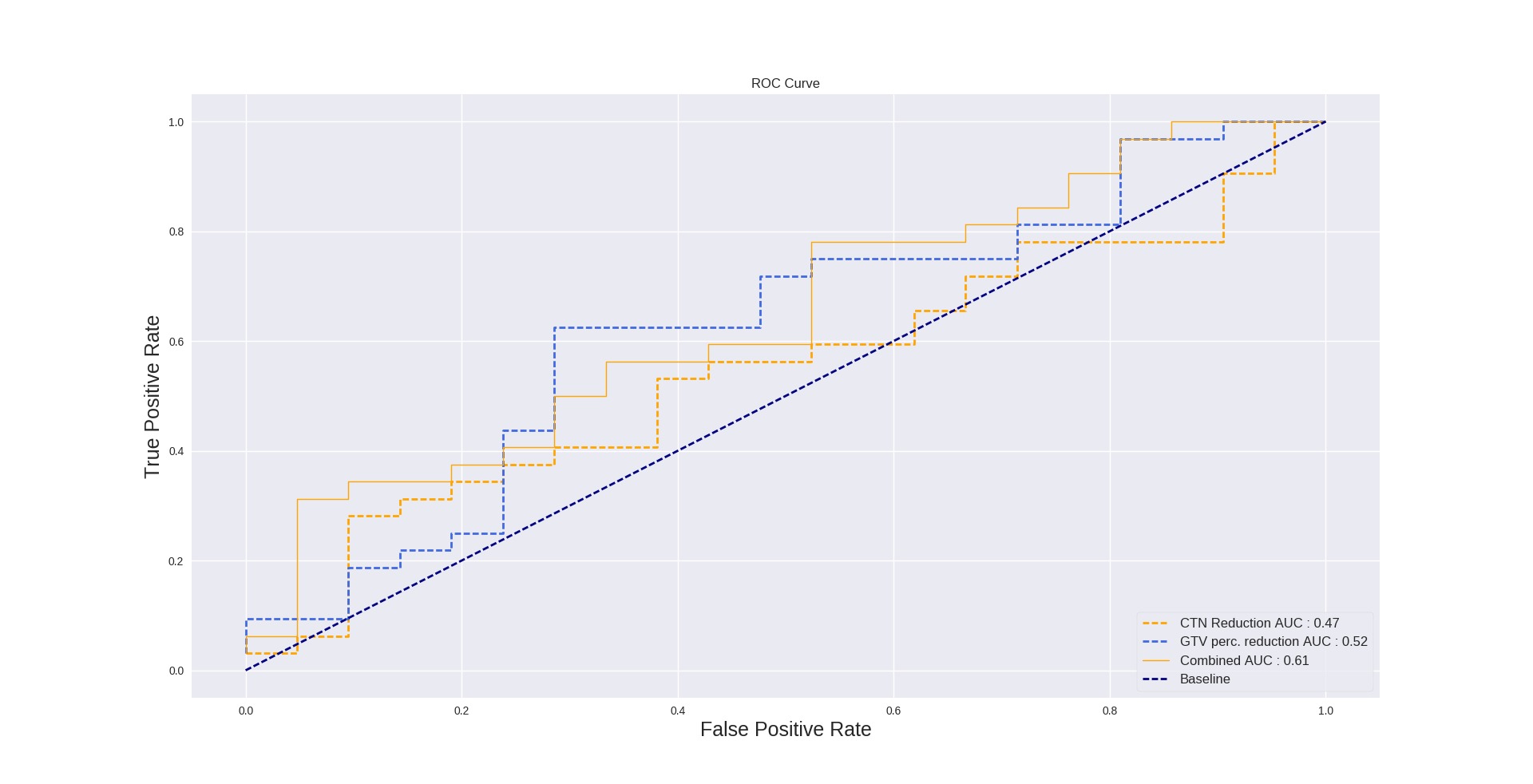}
  \label{fig:ROC}}
  \subfloat[]{\includegraphics[width=0.49\textwidth]{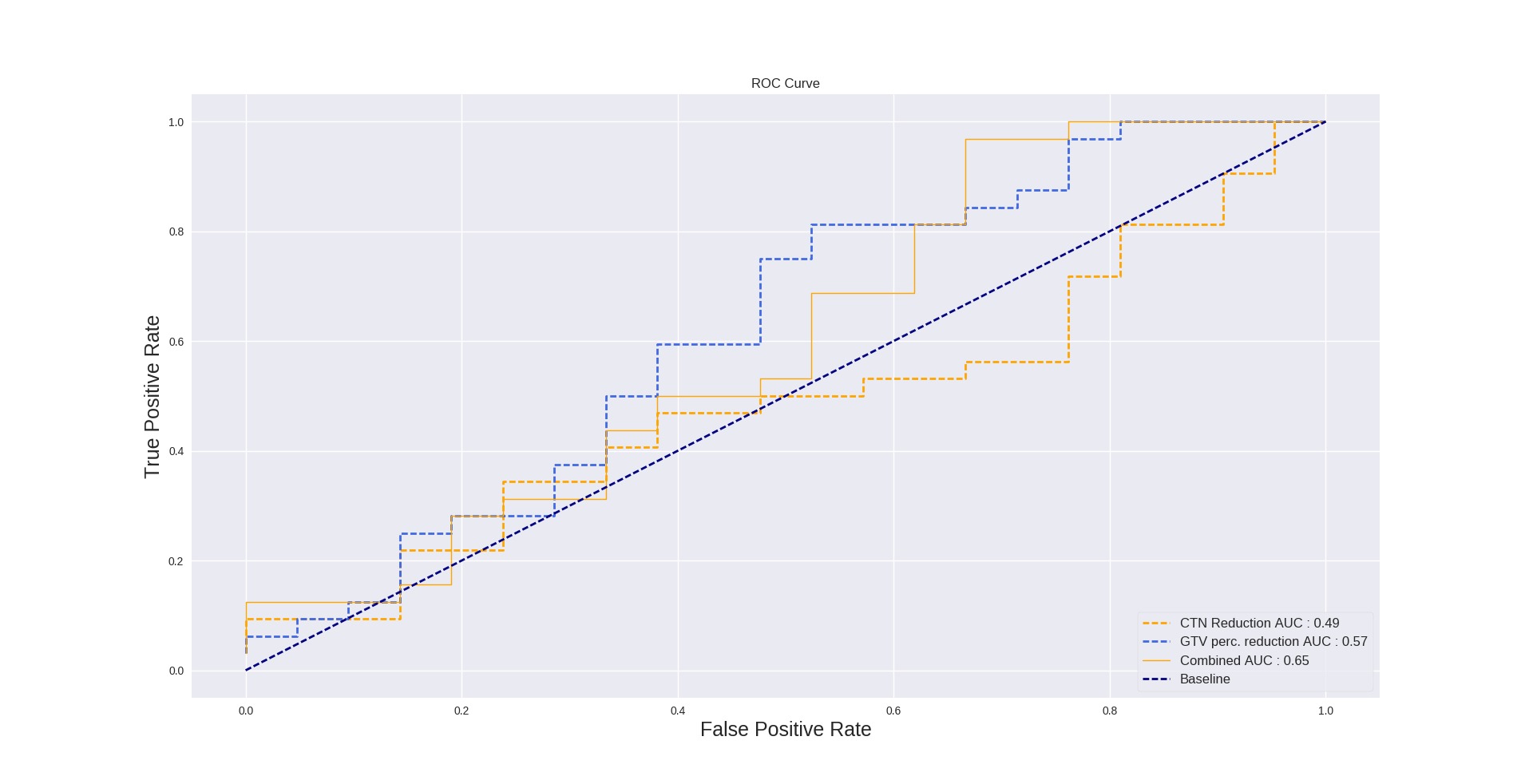}
  \label{fig:CTN_GTV_RANSAC}}
  \caption{The ROC curve for the three different cases is shown for (a) full data comprising of 6 weeks and, (b) 3 weeks data for early prediction.}
  \label{fig:ROC_RANSAC}
  \end{figure*}

  It can be observed that the AUC is better when using both the quantities as shown in Fig.~\ref{fig:ROC_RANSAC}.
  Also, early prediction using three-weeks data was found to give a better result than full six-weeks data.
  When fitted with the random sample consensus (RANSAC) algorithm, the correlation between CTN and GTV was found to be equal to 0.019.

  \subsubsection{Using CTN, GTV and Jacobian}
  Using the classifier based on CTN and GTV for logistic regression with the maximum AUC, we combined it with the proposed hypothesis based on Jacobians.
  The final classification was considered as consistent if both the classifiers agreed on their prediction.
  Whenever there is no mutual agreement, the prediction cannot be concluded.
  The results of using both classification methods are as follows:
  
  In the six-weeks data, the classifier trained using CTN, GTV using the full 6 weeks data was combined with the Jacobian method applied on the full 6 weeks data.
  For the three-weeks data also, a similar approach was followed.
  \begin{itemize}
  \item All 6 weeks: The sensitivity, specificity, precision, and accuracy were found to be 88\%, 40\%, 71\%, and 70\%, respectively, with a p-value of 0.15, implying the result is not statistically significant.
  15 out of the 32 responsive patients were predicted as responsive, while for 2 patients it was predicted as non-responsive. For the remaining 15 responsive patients, the classifier's prediction was not conclusive.
  4 out of the 21 non-responsive patients were predicted as non-responsive, while for 6 patients it was predicted as responsive. In the remaining 11 non-responsive patients the classifier did not give any conclusion.
  \item First 3 weeks: The sensitivity, specificity, precision, and accuracy were found to be 95\%, 50\%, 77\%, and 79\%, respectively, with a p-value of 0.0.
  20 out of the 32 responsive patients were predicted as responsive, while for 1 patient it was predicted as non-responsive. There was no conclusion in 11 of the responsive patients.
  6 out of the 21 non-responsive patients were predicted as non-responsive, while for 6 patients it was predicted as responsive. For 9 patients the classifier did not give any conclusion.
  \end{itemize}

  \section{Discussion}
  
  On our dataset, we applied the method for prediction using the features CTN and GTV reduction \cite{wen2017value} in the supervised logistic regression framework.
  We also presented our proposed unsupervised method using a hypothesis based on mean Jacobian in different regions of the GTV.
  Due to the lower amount of data, unsupervised methods are suitable as they do not have any dependency on the training sample size and quantity of data.
  Any supervised method requires cross-validation when the amount of data available is small.
  We observed that the training and validation loss is quite different on different folds.
  This suggests, that there could be possible overfitting.
  Fifty sample runs of the three-fold cross-validation model were executed to study the variation in the result while shuffling the data.
  When we trained the logistic regression model, it was found that there was considerable variation in the p-value.
  Finally, the classifier with the best AUC that was statistically significant ($p < 0.1$) was considered.
  
  Both (CTN/GTV and Jacobian) the above methods required ground truth delineation in the form of GTV contours.
  Table.~\ref{tab:summary} compares the result in all different cases.
  Both the methods, are able to perform early prediction using only the first three weeks of data.
  However, when used together, they seem to have a less mutual agreement regarding responsiveness and non-responsiveness of a patient.
  This is due to the observation that, 26 of the patients' prediction was not conclusive in the full data case, while for 20 patients it was inconclusive in the early prediction case.
  The method by Wen~\cite{wen2017value} has low specificity (33.3\%), while the Jacobian shows considerable improvement in the specificity (67\%).
  Higher specificity is always required in clinical applications.

  \section{Conclusion}
  CTN, GTV, and Jacobian are useful measures to classify responsive and non-responsive patients.
  Early prediction of treatment response at 3 months post-therapy is possible early by using only the first three weeks on-treatment CBCT data.
  However, more prospective data with a clinical trial is required to establish a classification method like ours, in clinical practice.

  \section*{Additional information}
  
  Competing Interests: The authors declare that they have no competing interests.
  
  \section*{Acknowledgment}
  The funding sources had no role in the study design, data collection, analysis of interpretation, or the writing of this manuscript.  

 \bibliographystyle{unsrt}
 \bibliography{references}

\end{document}